\documentclass[useAMS,usenatbib,referee]{mn2e}
\usepackage{graphicx,amssymb, txfonts}

\title[A correlation between the stellar and ${\rm [Fe\,II]}$ velocity dispersions]{A correlation between the stellar and ${\rm [Fe\,II]}$ velocity dispersions in Active Galaxies}

\author[Riffel et al.]{Rogemar. A. Riffel$^{1}$\thanks{E-mail:
rogemar@ufsm.br}, Thaisa Storchi-Bergmann$^{2}$, Rog\'erio Riffel$^{2}$, Miriani G. Pastoriza$^{2}$,
\newauthor   Alberto Rodr\'iguez-Ardila$^{3}$, Oli L. Dors Jr$^{4}$, Jaciara Fuchs$^{1}$, Marlon R. Diniz$^{1}$, 
\newauthor  Sch\"onell, A. J. J\'unior$^{1}$, Moir\'e G. Hennig$^{1}$, Carine Brum$^{1}$ \\
$^{1}$ Universidade Federal de Santa Maria, Departamento de F\'\i sica/CCNE, 97105-900, Santa Maria, RS, Brazil\\
$^{2}$ Universidade Federal do Rio Grande do Sul, Instituto de F\'\i sica, CP 15051, Porto Alegre 91501-970, RS, Brazil.\\
$^{3}$ Laborat\'orio Nacional de Astrof\'isica/MCT, Rua dos Estados Unidos 154, Itajub\'a, MG, Brazil.\\
$^{4}$ Universidade do Vale do Para\'iba, Av. Shishima Hifumi 2911,12244-000, São Jos\'e dos Campos SP, Brazil.
}
\begin{document}

%\date{Accepted 1988 December 15. Received 1988 December 14; in original form 1988 October 11}

\pagerange{\pageref{firstpage}--\pageref{lastpage}} \pubyear{2011}

\maketitle

\label{firstpage}

\begin{abstract}

We use near-infrared spectroscopic data from the inner few hundred parsecs of a sample of 47 active galaxies to investigate possible correlations between the stellar velocity dispersion ($\sigma_\star$), obtained from the fit of the K-band CO stellar absorption bands, and the gas velocity dispersion ($\sigma$) obtained from the fit of the emission-line profiles of [S\,{\sc iii}]$\lambda0.953\mu$m, [Fe\,{\sc ii}]$\lambda1.257\mu$m,  [Fe\,{\sc ii}]$\lambda1.644\mu$m and H$_2\lambda2.122\mu$m. While no correlations with $\sigma_\star$ were found for H$_2$ and [S\,{\sc iii}], a good correlation was found for the two [Fe\,II] emission lines, expressed by the linear fit  $\sigma_\star = 95.4\pm16.1 + (0.25\pm0.08)\times \sigma_{\rm  [Fe\,II]}$.  Excluding barred objects from the sample a better correlation is found between $\sigma_\star$ and $\sigma_{\rm  [Fe\,II]}$, with a correlation coefficient of $R=0.80$ and fitted by the following relation: $\sigma_\star = 57.9\pm23.5 + (0.42\pm0.10)\times \sigma_{\rm  [Fe\,II]}$. This correlation can be used to estimate $\sigma_\star$  in cases it cannot be  directly measured and the [Fe\,{\sc ii}] emission lines are present in the spectra, allowing to obtain the mass of the supermassive  black hole (SMBH)  from the $M_\bullet-\sigma_\star$  relation.  The scatter from a one-to-one relationship between $\sigma_\star$ and its value derived from $\sigma_{\rm [Fe\,II]}$ using the equation above for our sample is 0.07\,dex, which is smaller than that obtained in previous studies which use $\sigma_{\rm [O\,III]}$ in the optical as a proxy for $\sigma_\star$. The use of $\sigma_{\rm [Fe\,II]}$ in the near-IR instead of $\sigma_{\rm [O\,III]}$ in the optical is a valuable option for cases in which optical spectra are not available or are obscured, as is the case of many AGN. The comparison between the SMBH masses obtained using the  $M_\bullet-\sigma_\star$ relation in which $\sigma_\star$ was directly measured with those derived from $\sigma_{\rm [Fe\,II]}$ reveals only a small average difference of  $\Delta{\rm log} M_\bullet =0.02$ with a scatter of 0.32\,dex for the complete sample and  $\Delta{\rm log} M_\bullet =0.00$ with a scatter of 0.28\,dex for a sub-sample excluding barred galaxies.

\end{abstract}

\begin{keywords}
galaxies: active -- galaxies: nuclei -- infrared: galaxies -- black holes
\end{keywords}

\section{Introduction} \label{intro}

In the present paradigm of galaxy evolution, most galaxies which form a bulge also form a supermassive black hole (SMBH) in their nuclei \citep[e.g.][]{magorrian98,richstone98,ferrarese00,gebhardt00}. The central SMBH seems to play a fundamental role in the galaxy evolution and cosmological simulations without considering the presence of a SMBH and its associated feedback predict masses for the galaxies much higher than those observed  \citep{dimateo05,springel05,bower06}. In a scenario of co-evolution of the SMBH and its host galaxy, mass accretion to the central region of the galaxy leads to the growth of the galaxy bulge, while the feeding of the SMBH triggers episodes of nuclear activity which results in feedback in the form of radiation pressure and mass ejections from the accretion disk surrounding the SMBH. This episodic feedback may halt the mass accretion to the galaxy preventing its growth in the active phase \citep{nemmen07}. This co-evolution may be the mechanism which leads to the empirical relation between the mass of the SMBH and the stellar velocity dispersion of the bulge $M_\bullet-\sigma_\star$, [\citet{ferrarese05}; but see also \citet{jahnke11}].

The $M_\bullet-\sigma_\star$ relation has been extensively used to estimate the mass of SMBHs  from the stellar kinematics, as direct determinations for the SMBH masses only are possible for the closest galaxies for which  the radius of influence of the SMBH can be resolved \citep[e.g][]{ferrarese05}. Although allowing to estimate the masses of the SMBHs for a large number of galaxies, the use of the $M_\bullet-\sigma_\star$ relation requires the measurement of $\sigma_\star$, which is not always easy to obtain, particularly in active galaxies, where the AGN continuum dilutes the stellar absorption lines. In order to overcome this difficulty, a number of scaling relations using the width and luminosities of emission lines to determine $M_\bullet$ have been proposed \citep[e.g.][]{booth11,wu09,peterson08,salviander06,vestergaard06,greene06,greene05,kaspi05,onken04,nelson96}. Nevertheless, most of these relations are for the optical domain of the electromagnetic spectrum. With the improvement of infrared (IR) detectors, IR spectra of many AGNs have become recently available, and have the advantage of being less affected by reddening than optical spectra. In the present paper, we investigate the possibility of using the widths of emission lines in the near-IR as proxies for $\sigma_\star$. 

Recent studies by our group, using near-infrared (hereafter near-IR) integral field spectroscopy of active galaxies, have allowed the mapping of the flux distributions and kinematics of the molecular (H$_2$) and ionized gas. We have found, in particular, that the H$_2$ usually shows small velocity dispersions and a velocity field dominated by rotation, while the ionized gas shows higher velocity dispersions and is  not dominated by rotation  \citep[e.g.][]{n4051,n7582,mrk1066-kin,mrk1157}. The kinematics and flux distributions are also consistent with a location for the H$_2$ gas in the galaxy plane, while the ionized gas, and, in particular [Fe\,{\sc ii}] extends to high galactic latitudes.
%This result suggests that H$_2$ $\sigma$ could be a proxy for the stellar velocity dispersion  and could thus be used to estimate SMBH masses ($M_\bullet$) using the $M_\bullet-\sigma_\star$ relation \citep[e.g.][]{ferrarese00,gebhardt00,tremaine02,graham11}. In the other hand, the velocity dispersion for the [Fe\,{\sc ii}] emitting gas is systematically larger than those observed for the stars. 

In the present paper we investigate the correlation between the gas and stellar kinematics derived from near-IR spectroscopy, with the goal of looking for a ''proxy'' for $\sigma_\star$ among the brightest emission lines in this wavelength range. This paper is organized as follows: in Section~2, we describe the sample and the observational data; in Section~3  we describe the methods used to measure the stellar and gaseous velocity dispersion. The results are presented in Section~4 and discussed in Section~5, while the conclusions are shown in Section~6.

\section{Observational data}

The spectroscopic data used in this work are from \citet{atlas,ardila05} and \citet{ardila04}. The sample comprises 47 active galaxies with a range of activity types, and the spectra cover, on average, the inner 300\,pc radius of the galaxies.

The near-infrared spectra were obtained with the NASA 3\,m Infrared Telescope Facility (IRTF), using the SpeX spectrograph in the short cross-dispersed mode (SXD, 0.8-2.4$\mu$m). The detector employed was a
1024$\times$1024 ALADDIN 3 InSb array with a spatial scale of 0.15"/pixel. A 0.8''$\times$15'' slit was used and the spectral resolution is 300~km\,s$^{-1}$, obtained from the measurement of the full width at half maximum (FWHM) of  Arc lamp lines, or 127~km\,s$^{-1}$ in $\sigma$.  The data reduction followed standard procedures. For more details on the instrumental configuration, data reduction, calibration processes and details of the extraction of the spectra  see \citet{atlas}. 

The above sample was chosen for this work because it is an unique dataset of near-IR spectroscopy of active galaxies, observed with the same instrumental setup (thus with no instrumental bias), covering the near-IR Z, J, H and K bands, and including several emission and absorption features allowing the comparison of the stellar and gas kinematics.

\section{Methods}

In order to obtain the gaseous velocity dispersion $\sigma$, we fitted the emission-line profiles of [S\,{\sc iii}]$\lambda0.953\mu$m, [Fe\,{\sc ii}]$\lambda1.257\mu$m, [Fe\,{\sc ii}]$\lambda1.644\mu$m and H$_2\lambda2.122\mu$m by single Gaussian curves and adopted as the measured velocity dispersion the $\sigma$ of the Gaussian. These emission lines have been chosen because they are  the strongest  in the near-IR spectra of active galaxies \citep[e.g.][]{atlas}. We excluded the H and He recombination lines due to uncertainties in the fit for type 1 objects, for which is not always easy to separate the narrow from the broad components. The fitting of the emission-line profiles was done by adapting the {\sc profit} routine \citep{profit}, which outputs  the emission-line flux, the centroid velocity,  the velocity dispersion and the uncertainties for each of these parameters. The velocity dispersion was then corrected for the instrumental $\sigma_{inst}=$127~km\,s$^{-1}$, which was subtracted in quadrature from the $\sigma$ obtained from the fit of the Gaussians to the line profiles.

We measured the stellar velocity dispersion ($\sigma_\star$) using the penalized Pixel-Fitting (pPXF) method  of \citet{ppxf} in order to fit the CO absorptions bands at $\sim2.3\mu$m  in the K-band. The pPXF method requires the use of stellar spectra as templates. We used for this the spectra of 60 late-type stars, 40 of them from the Gemini Near-IR Late-type stellar library \citep{winge09} and the remaining 20 spectra are from stars with public NIFS observations in the Gemini data archive (Diniz et al., 2012 in preparation). The uncertainties on the measurements of $\sigma_\star$ were estimated using 100 Monte-Carlo iterations as in \citet{mrk1157}. 

\begin{figure*}
\centering
\begin{minipage}{1\linewidth}
%\begin{minipage}{0.45\linewidth}
\includegraphics[scale=0.25]{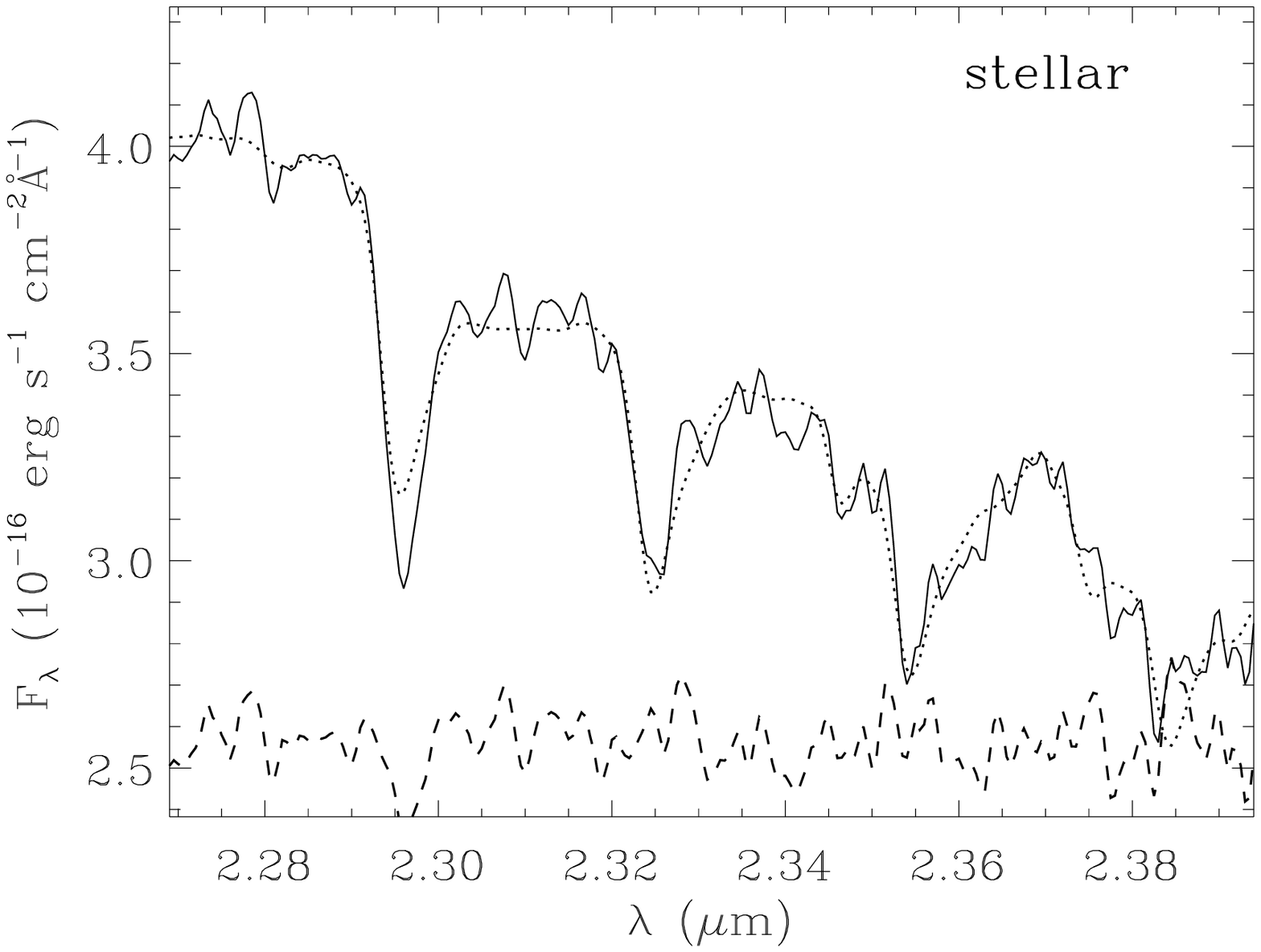}
\includegraphics[scale=0.25]{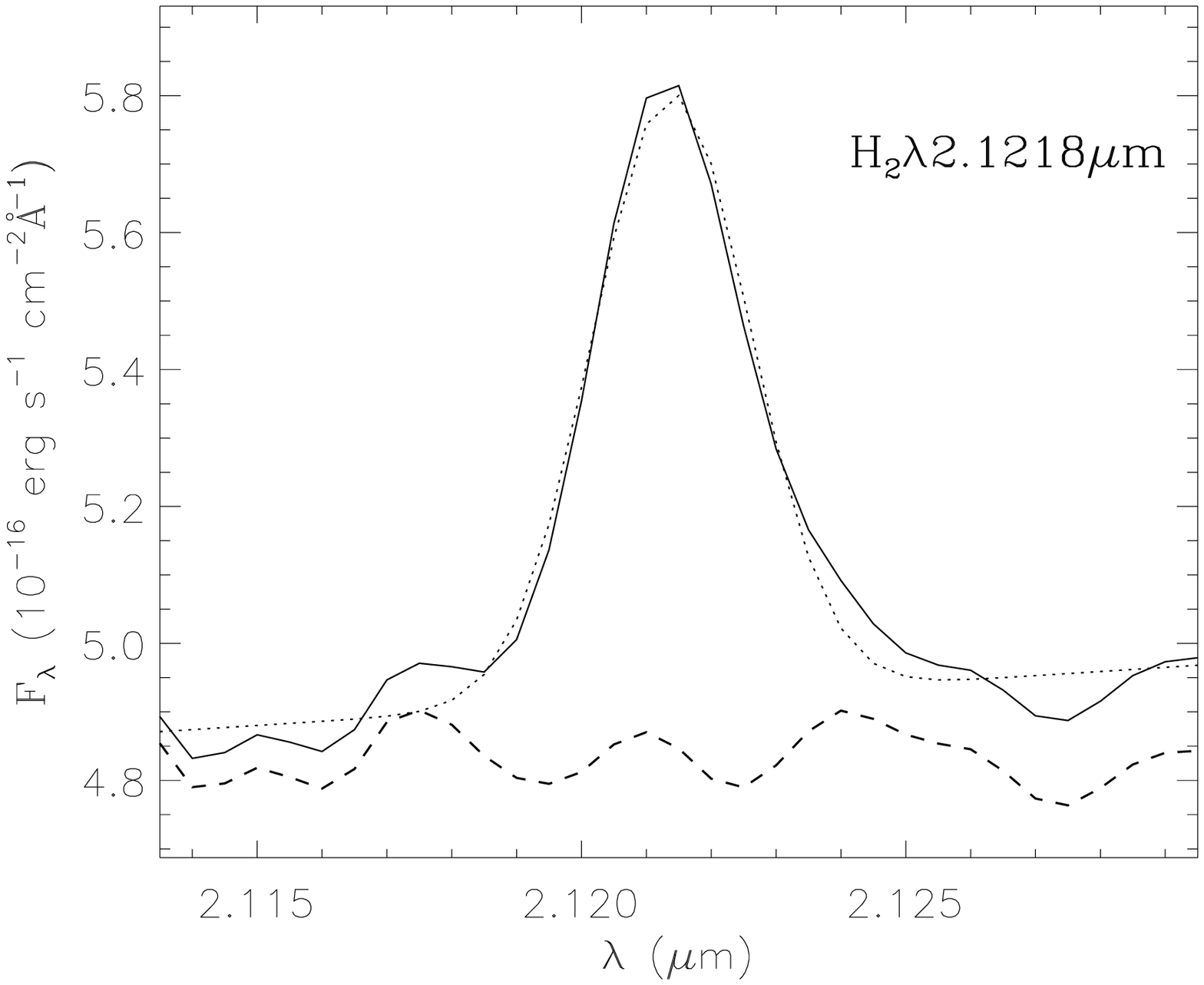}
\end{minipage}
\begin{minipage}{1\linewidth}
%\begin{minipage}{0.45\linewidth}
\includegraphics[scale=0.25]{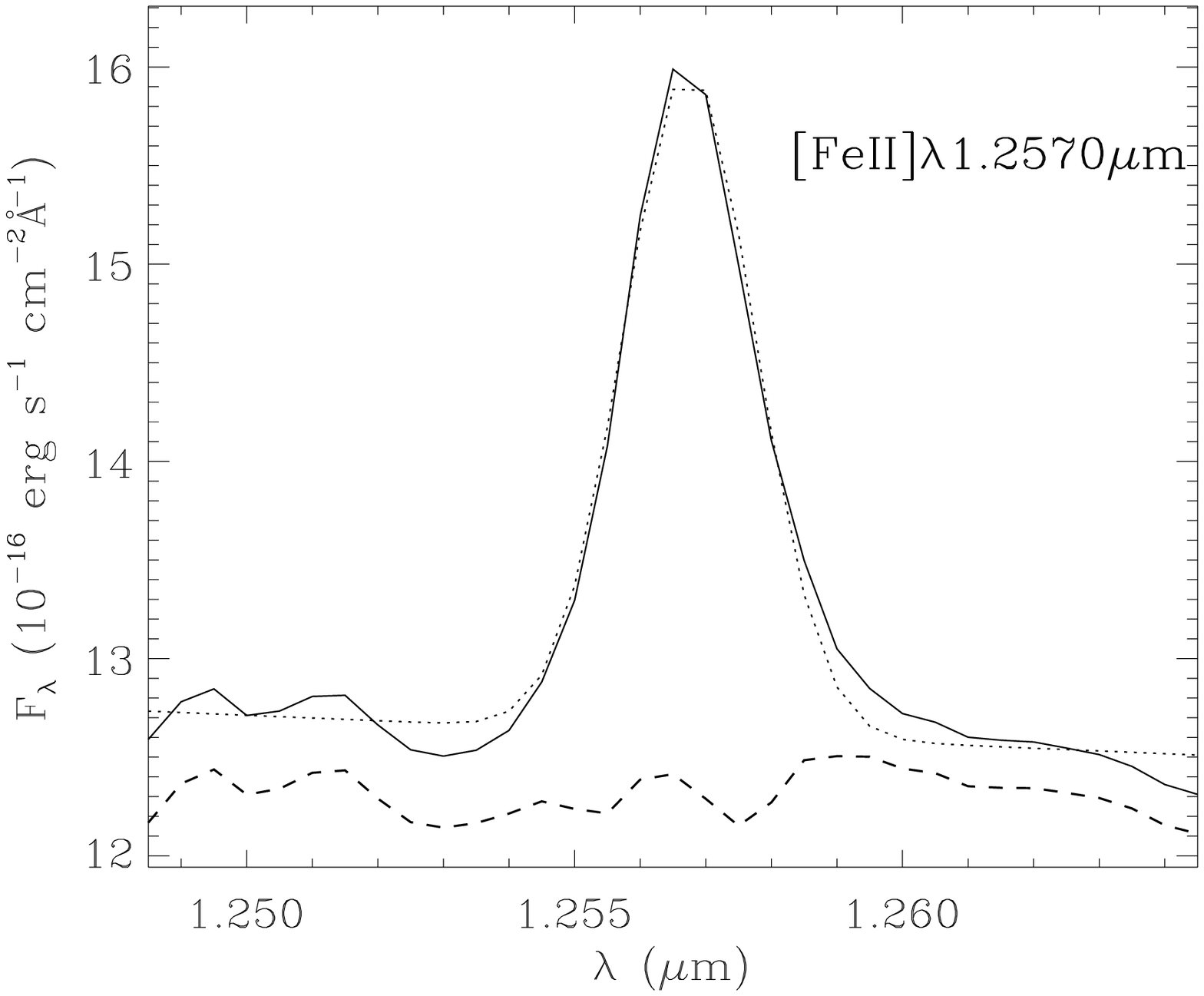}
\includegraphics[scale=0.25]{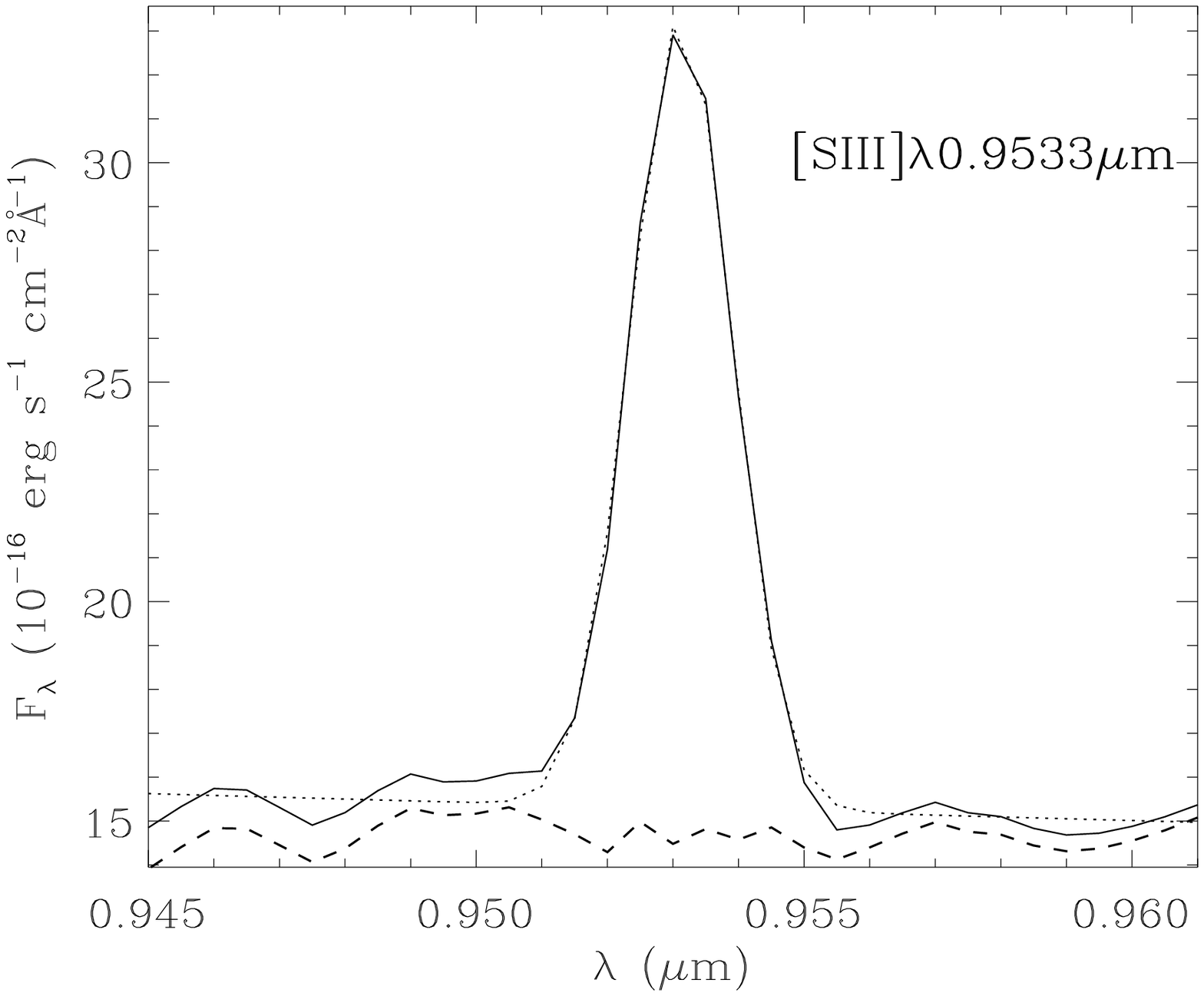}
\end{minipage}
\caption{Sample fits of the spectrum of the galaxy  NGC\,5929. Top left: fit of the stellar absorption spectra to obtain the stellar kinematics.  Remaining panels: fit of the emission-line profiles of H$_2$, [Fe\,{\sc ii}] and [S\,{\sc iii}]. The observed spectra are shown as continuous lines, the fits as dotted lines and the residuals as dashed lines.}
\label{fit}  
\end{figure*}

In order to illustrate our procedures, we show in Fig.~\ref{fit} sample fits of the CO absorption band heads at 2.3\,$\mu$m using the {\sc ppxf} method, as well as fits of the emission-line profiles using {\sc profit} for the spectrum of the galaxy NGC5929.

\section{Results}
The resulting measurements for the stellar and gas velocity dispersions for the galaxies of our sample are shown in Table~\ref{table}. The dashes in the Table are due to the fact that for a few objects we were not able to measure one or more values due to the absence of the absorption/emission lines or due to a low signal-to-noise ratio.

\begin{table*}
\caption{Stellar and gas velocity dispersions for the galaxies of the sample.}
\label{table}
\centering
\begin{tabular}{l l l cccc}
\hline
Object      & Hubble Type$^a$ & Nuclear Activity$^a$ & $\sigma_\star$ (km s$^{-1}$)  &   $\sigma_{\rm H_2}$ (km s$^{-1})$  &  $\sigma_{\rm  [Fe\,II]}$ (km s$^{-1}$) &$\sigma_{\rm [SIII]}$ (km s$^{-1}$)  \\
\hline       
MRK334	    & Sbc    	  & Sy1  &$<$127      & $<$127      & 180$\pm$9   & 278$\pm$14	\\
NGC34	    & S0/a    	  & Sy2  &160$\pm$23  & 198$\pm$17  & 227$\pm$19  & 215$\pm$18	\\
NGC262	    & S0/a  	  & Sy2  &--	      & $<$127      & 263$\pm$17  & 245$\pm$16	\\
MRK993	    & Sa     	  & Sy2  &132$\pm$14  & 185$\pm$10  & 214$\pm$12  & 287$\pm$16	\\
NGC591	    & SB0/a 	  & Sy2  &130$\pm$21  & $<$127      & 274$\pm$17  & 229$\pm$14	\\
MRK573	    & SB0   	  & Sy2  &$<$127      & $<$127      & 172$\pm$12  & 211$\pm$15	\\
NGC1097     & SB(s)b 	  & Sy1  &165$\pm$13  & $<$127      & 166$\pm$12  & 381$\pm$28	\\
NGC1144     & E           & Sy2  &206$\pm$40  & 175$\pm$13  & 230$\pm$18  & 233$\pm$18	\\
MRK1066     & S0/a        & Sy2  &$<$127      & $<$127      & 206$\pm$15  & 234$\pm$17	\\
NGC1275     & S0	  & Sy2  &--	      & 173$\pm$21  & 313$\pm$38  & 564$\pm$68	\\
NGC1614     & SB(s)c      & Sb   &$<$127      & $<$127      & 203$\pm$12  & 217$\pm$13	\\
MCG-5-13-17 & SB0/a 	  & Sy1  &162$\pm$11  & 191$\pm$22  & 162$\pm$19  & 201$\pm$23	\\
NGC2110     & E-S0        & Sy2  &184$\pm$16  & $<$127      & 248$\pm$17  & 305$\pm$21	\\
ESO428-G014 & S0          & Sy2  &$<$127      & 161$\pm$11  & 233$\pm$18  & 253$\pm$18	\\
MRK1210     & S?	  & Sy2	 &181$\pm$24  & $<$127      & 316$\pm$32  & 342$\pm$35	\\
MRK124	    & S? 	  & NLS1 &222$\pm$23  & 235$\pm$27  & 335$\pm$39  & 244$\pm$29	\\
MRK1239     & E-S0	  & NLS1 &--	      & --	    & $<$127      & 370$\pm$35	\\
NGC3227     & SB(s)a      & Sy1	 &128$\pm$3   & 157$\pm$24  & 314$\pm$48  & 276$\pm$42	\\
H1143-192   & -- 	  & Sy1  &--	      & --	    & 170$\pm$28  & 410$\pm$68	\\
NGC3310     & SB(r)bc     & Sb	 &142$\pm$18  & $<$127      & 130$\pm$9   & 153$\pm$10	\\
PG1126-041  & S 	  & QSO  &--	      & $<$127      & 254$\pm$34  & 363$\pm$48	\\
NGC4051     & SB(rs)bc    & NLS1 &$<$127      & $<$127      & 142$\pm$6   & 215$\pm$9	\\
NGC4151     & SB(rs)ab    & Sy1	 &136$\pm$19  & $<$127      & 194$\pm$41  & 231$\pm$48	\\
MRK766	    & SB(s)a	  & NLS1 &$<$127      & $<$127      & 133$\pm$20  & 188$\pm$19	\\
NGC4748     & S?	  & NLS1 &$<$127      & 170$\pm$15  & 171$\pm$21  & 273$\pm$23	\\
TONS0156    & --	  & QSO	 &--	      & --	    & $<$127      & 712$\pm$200	\\
MRK279	    & S0	  & NLS1 &138$\pm$12  & $<$127      & 209$\pm$49  & 335$\pm$79	\\
NGC5548     & S0/a        & Sy1	 &$<$127      & $<$127      & 165$\pm$43  & 213$\pm$55	\\
MRK478	    & Sc	  & NLS1 &--	      & 161$\pm$22  & 171$\pm$23  & 390$\pm$53	\\
NGC5728     & SBa         & Sy2	 &$<$127      & 132$\pm$23  & $<$127      & 223$\pm$38	\\
PG1448+273  & E?	  & QSO  &--	      & $<$127      & --  	  & 303$\pm$1	\\
MRK684	    & Sab	  & Sy1  &170$\pm$17  & 351$\pm$18  & --  	  & 540$\pm$27	\\
MRK291	    & SBa	  & NLS1 &--	      & 129$\pm$14  & 144$\pm$16  & 167$\pm$18	\\
MRK493	    & SBb         & NLS1 &--	      & $<$127      & 232$\pm$12  & 412$\pm$21	\\
NGC5929     & Sa          & Sy2	 &158$\pm$23  & $<$127      & 208$\pm$26  & 181$\pm$23	\\
NGC5953     & S0/a        & Sy2	 &149$\pm$4   & 187$\pm$8   & 289$\pm$13  & 289$\pm$13	\\
PG1612+261  & --	  & QSO  &166$\pm$18  & --	    & 266$\pm$53  & 262$\pm$52	\\
MRK504	    & S?	  & NLS1 &$<$127      & $<$127      & --  	  & 333$\pm$38	\\
3C351	    & --	  & BLRG &--	      & --	    & --    	  & 187$\pm$14	\\
ARP102B     & E0	  & Sy1	 &$<$127      & 155$\pm$9   & 195$\pm$10  & 299$\pm$16	\\
1H 1934-063 & --	  & NLS1 &184$\pm$9   & 191$\pm$8   & 238$\pm$11  & 291$\pm$13	\\
MRK509	    & E-S?	  & Sy1	 &$<$127      & --	    & 159$\pm$39  & 370$\pm$91	\\
1H2107-097  & --	  & Sy1	 &$<$127      & --	    & $<$127      & 256$\pm$56	\\
ARK564	    & SBc	  & NLS1 &--	      & 180$\pm$10  & 176$\pm$10  & 260$\pm$15	\\
NGC7469     & SBbc        & Sy1  &--	      & --	    & 161$\pm$15  & 283$\pm$18  \\
NGC7682     & SBab        & Sy2	 &183$\pm$40  & $<$127      & 198$\pm$27  & 205$\pm$28	\\
NGC7714     & SB(s)b      & HII	 &$<$127      & $<$127      & 165$\pm$13  & 184$\pm$14	\\
\hline
\multicolumn{7}{|l|}{$^a$ The Hubble type and Nuclear Activity were taken NASA/IPAC Extragalactic Database (NED) and Hyperleda Database \citep{paturel03}.} \\
\end{tabular} 
\end{table*}

\begin{figure}
 \centering
 \includegraphics[scale=0.55]{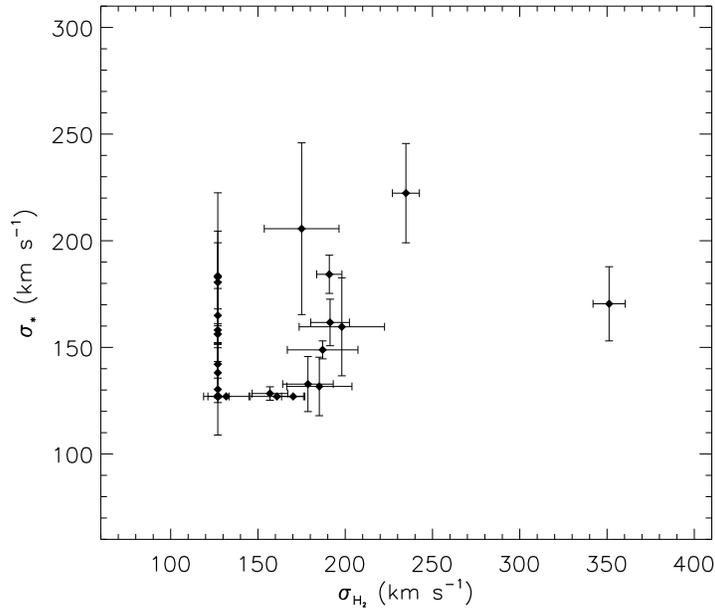}
\caption{Comparison between the gas velocity dispersions obtained from the $\rm H_{2}\lambda2.122\mu$m emission line ($\sigma_{\rm H_{2}}$) and the stellar velocity dispersions from the CO stellar absorptions at  $\thicksim$2.3$\mu$m ($\sigma_\star$). Points with no error bars in one or both axes represent measurements that are unresolved by our observations and should be considered as upper limits.} 
 \label{h2}  
 \end{figure}

\begin{figure}
 \centering
 \includegraphics[scale=0.55]{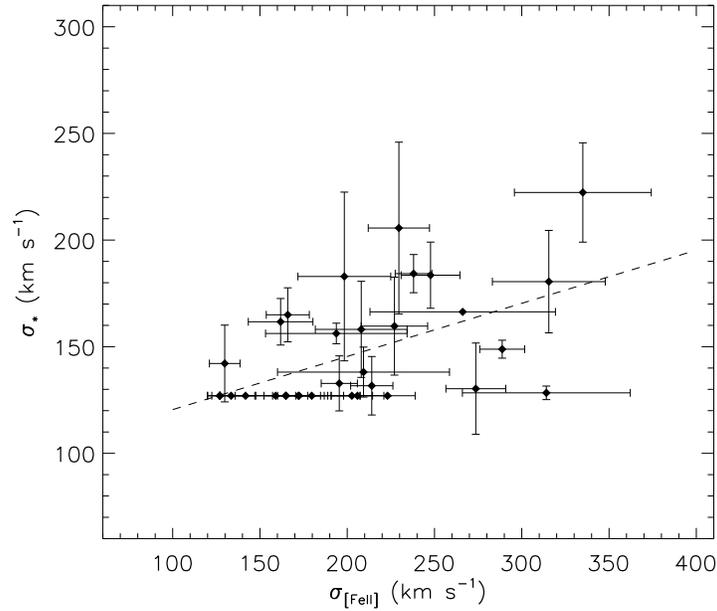}
 \caption{As Fig.~\ref{h2} but for gas velocity dispersion  derived  using [Fe\,{\sc ii}]$\lambda1.257\mu$m. The dashed line represents the best linear fit of the data, given by: 
$\sigma_\star = 95.4\pm16.1 + (0.25\pm0.08)\times \sigma_{\rm [Fe\,II]}$.
} 
 \label{fe}  
 \end{figure}

\begin{figure}
 \centering
 \includegraphics[scale=0.55]{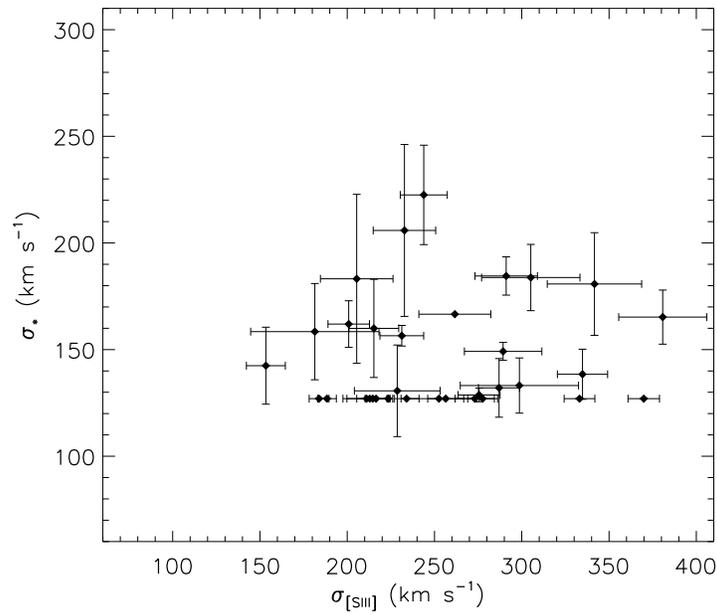}
\caption{As Fig.~\ref{h2} but for gas velocity dispersion  derived  using [S\,{\sc iii}]$\lambda 0.953\mu$m.} 
 \label{siii}  
 \end{figure}

\begin{figure}
 \centering
 \includegraphics[scale=0.6,angle=-90]{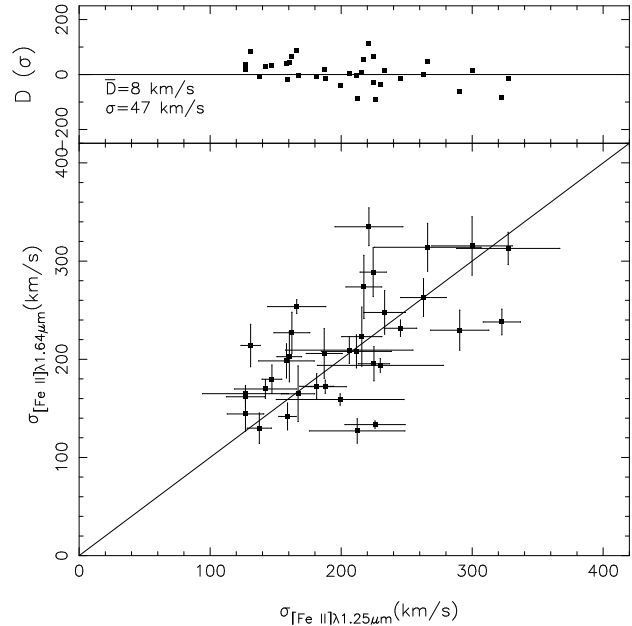}
\caption{$\sigma_{\rm [Fe\,II]\lambda1.644}$ vs. $\sigma_{\rm [Fe\,II]\lambda1.257}$. The solid line represents a one-to-one relationship. The top panel shows the difference between the values of $\sigma_{\rm  [Fe\,II]\lambda1.644}$ and $\sigma_{\rm  [Fe\,II]\lambda1.257}$ defined as $D(\sigma)=\sigma_{\rm  [Fe\,II]\lambda1.64 i}-\sigma_{\rm  [Fe\,II]\lambda1.25 i}$, as well as the average value of this difference ($\bar{D}$).
} 
 \label{fefe}  
 \end{figure}

We have  looked for correlations between the stellar and the gaseous velocity dispersions using the values of Table~\ref{table} to construct  the graphs of Figures~\ref{fe}, \ref{h2} and \ref{siii}. We have used the IDL routine {\sc r$_-$correlate} to obtain the Spearman correlation coefficient $R$ for each graph. Figure~\ref{h2} shows $\sigma_\star$ vs. $\sigma_{\rm H_2}$. The range of the $\sigma_\star$ and $\sigma_{\rm H_2}$ values is approximately the same, something we have noticed in our previous studies of individual galaxies using integral field spectroscopy of the inner hundreds of parsecs \citep[e.g.][]{n4051,n7582,mrk1066-kin,mrk1157}. Nevertheless, we have obtained only a very weak correlation between $\sigma_{\rm H_2}$ and $\sigma_\star$, with $R=0.35$, but we note that the H$_2$ line is unresolved for several objects.

% The  continuous line shows a one-to-one relationship between $\sigma_\star$ and $\sigma_{\rm H_2}$. The top panel of Fig.~\ref{h2}  shows the difference between the values of $\sigma_{\rm H_2}$ and $\sigma_\star$ defined as $D(\sigma)=\sigma_{\rm H_2 i}-\sigma_{\star i}$, as well as the average value of this difference. Solid lines represent equality of the two values and the dashed line represents a linear fit to the data. 

The relation between $\sigma_\star$ and $\sigma_{\rm  [Fe\,II]}$ is presented in Figure~\ref{fe}, showing that $\sigma_{\rm  [Fe\,II]}$  is usually higher than $\sigma_\star$, which is also in agreement with the results from the integral field spectroscopic studies above. A better correlation is observed between  $\sigma_\star$ and $\sigma_{\rm  [Fe\,II]}$ than with $\sigma_{\rm H_2}$, corresponding to a correlation coefficient $R=0.56$ obtained as described above. We fitted the data by a linear equation of the form $\sigma_\star = A + B\times \sigma_{\rm  [Fe\,II]}$ using the IDL routine {\sc linmix$_-$err},  which
uses a Bayesian approach to linear regression with errors in both variables and takes into account upper limits for the measurementes \citep{kelly07}.  The best fit to the data is given by 
\begin{equation} \label{msigfe}
\sigma_\star = 95.4\pm16.1 + (0.25\pm0.08)\times \sigma_{\rm  [Fe\,II]}
\end{equation}
which is shown as a dashed line in Fig.~\ref{fe}. 

Finally, the relation between $\sigma_\star$ and $\sigma_{\rm [SIII]}$ is shown in Figure~\ref{siii}, resulting in a correlation coefficient $R=0.32$, suggesting only a very weak correlation. This figure also shows that $\sigma_{\rm [SIII]}$ is larger than $\sigma_\star$ by more then a hundred km\,s$^{-1}$, on average.

\section{Discussion}

The use of the velocity dispersion from the Narrow Line Region emission lines as a proxy for $\sigma_\star$ in order to obtain an estimate for the SMBH mass via the M$_{\rm BH}-\sigma_\star$ relation in active galaxies is not new. It has been previously used in the optical domain, where the emission line most commonly used is [O\,{\sc iii}]$\lambda5007$ \citep[e.g.][]{wu09,salviander06,kaspi05,onken04}. This emission line has been used instead of $\sigma_{\star}$ because in active galaxies $\sigma_\star$  cannot be easily measured due to dilution of the stellar absorption lines by the AGN continuum or its scattered light.

In this paper we present an alternative to be used in the near-IR. As shown above, we found a correlation between $\sigma_\star$ and $\sigma_{\rm [Fe\,II]}$, indicating that  the latter can be used to estimate $\sigma_\star$ using equation \ref{msigfe}.  [Fe\,{\sc ii}] has two similarly strong emission lines which can be observed in the near-IR:  [Fe\,{\sc ii}]$\lambda1.257$ in the J band and  [Fe\,{\sc ii}]$\lambda$1.644 in the H band. In Figure~\ref{fefe} we present a comparison between the $\sigma$ of these two lines, where the solid line shows an one-to-one relationship. This comparison shows that the width of these lines is the same within the errors, with a mean difference of $\sigma_{\rm  [Fe\,II]\lambda1.644}$-$\sigma_{\rm  [Fe\,II]\lambda1.257}=8$~km\,s$^{-1}$ and a scatter of 47~km\,s$^{-1}$, as seen in the top panel of Fig.\,\ref{fefe}. This scatter may be partially due to the fact that  [Fe\,{\sc ii}]$\lambda1.644$ is close in wavelength to Brackett\,12, which usually appears in absorption and may affect the measurement of [Fe\,{\sc ii}] line.

%Thus,  $\sigma_{ [Fe\,{\sc ii}]\lambda1.64}$ can also be used to estimate $\sigma_\star$ via equation~\ref{msigfe}. 

Why is $\sigma_{[Fe\,II]}$  better correlated with $\sigma_{\star}$ than $\sigma_{\rm H_2}$? As pointed out in the Introduction, our previous studies using Integral field spectroscopy  \citep{n4051,n7582,mrk1066-kin,mrk1157} showed that the H$_2$ kinematics frequently shows a rotation pattern and a smaller velocity dispersion than that of the  ionized gas. This also led to the conclusion that the H$_2$ gas was more restricted to the galaxy plane, while the ionized gas -- and in particular [Fe\,{\sc ii}] --  extended to higher galactic latitudes. The integrated value of $\sigma_{\star}$ from the nuclear region of galaxies is dominated by the contribution of bulge stars, which are not restricted to the plane, showing a ``hotter'' kinematics. Thus, it can be understood that the velocity dispersion of gas which is restricted to the plane does not correlate with that of bulge stars, while that of  gas extending to higher latitudes, similar to those of the bulge  -- such as the   [Fe\,{\sc ii}] emitting gas -- is correlated to that of the bulge stars. The higher values of $\sigma_{\rm [Fe\,II]}$ relative to $\sigma_{\star}$ are probably due to extra heating provided by a nuclear AGN outflow.

Our results can be compared with previous ones in the optical using the $\sigma$ of the  [O\,{\sc iii}]$\lambda$5007 emission line as a proxy for $\sigma_{\star}$. For a sample of 66 Seyfert galaxies, \citet{nelson96} found a scatter of 0.20\,dex around a one-to-one relation between $\sigma_\star$ and  $\sigma_{\rm [O\,III]}$, while \citet{onken04} found a smaller scatter of 0.15\,dex for a sample of 16 AGNs, for which 25\% of their sources have $\sigma_{\rm [O\,III]}$ deviating by more than 0.20\,dex from the values expected based on their $\sigma_\star$. We found a scatter of 0.07\,dex between the values obtained via equation~\ref{msigfe} and the measured values for $\sigma_\star$, which is thus smaller than that for $\sigma_{\rm [O\,III]}$. 

 A cautionary note is the observation of recent spatially resolved studies  (e.g. our previous studies already mentioned) that the [Fe\,{\sc ii}] emission originates at least in part in outflowing gas. Thus, the width of the line is not only due to orbital motion in the gravitational potential of the galaxy, but is also due to broadening by the outflow, what is consistent with the observation that $\sigma_{\rm [Fe\,II]}$ is larger than $\sigma_\star$. Similar outflows -- most probably the same -- are observed in the [O\,{\sc iii}] emitting gas \citep[e.g.][]{fischer11,fischer10,crenshaw10,crenshaw09,crenshaw07,das07,das05}. Nevertheless, this line has been frequently used as a proxy for $\sigma_{\star}$ as discussed above, due to the lack of a better indicator. Our argument is that the $\sigma_{\rm [Fe\,II]}$ is at least as good $\sigma_{\star}$ proxy as $\sigma_{\rm [O\,III]}$, and can be used when the latter is not available.

We thus propose the use of $\sigma_{\rm  [Fe\,II]}$ to obtain $\sigma_\star$ via eq.\,\ref{msigfe} in cases for which it is not possible to measure the stellar kinematics of the galaxy, and the optical spectrum is obscured or not available, so that  the [O\,III]$\lambda$5007 emission line is also not available.  Nevertheless, this suggestion should be used with care, since the $M_{\bullet}-\sigma_\star$ is calibrated from a parent sample of mostly early-type galaxies and, as seen in Table\,\ref{table} most of the objects of our sample are late-type. Late-type galaxies can have distinct $\sigma_\star$ values than early-type galaxies, since the orbits of the stars in a disk are different than the orbits of the stars in the bulge. Additionally, some of the galaxies of our sample have bars, circumnuclear star forming rings or nuclear starbursts or even are classified as peculiar objects and thus the $\sigma_\star$ measured for these objects could be very different than those for the classical bulge, used to calibrate the $M_{\bullet*}-\sigma_\star$ relationship. 

\subsection{The effect of galaxy morphology on the $M_\bullet - \sigma$ relation}

%As pointed out in the Introduction, previous near-IR integral field spectroscopic studies by our group have shown that the emission lines of molecular hydrogen have smaller $\sigma$ values than those of ionized gas for active galaxies, with the range of $\sigma$ values comparable to those of $\sigma_{\star}$ \citep{n4051,n7582,mrk1066-kin,mrk1157}. This result suggests that  $\sigma_{\rm H_2}$ could be used as a proxy for $\sigma_{\star}$ in cases the latter is not available. The results presented above for a larger sample of 47 galaxies nevertheless show that $\sigma_{\rm H_2}$  does not present a good correlation with $\sigma_\star$ and thus is not very useful to estimate $M_\bullet$. We have also shown that the width of [S\,{\sc iii}]$\lambda0.95332\mu$m  -- another strong emission line in most AGNs -- presents also only a very weak correlation with the $\sigma_{\star}$.  On the other hand, we have shown that $\sigma_{ [Fe\,{\sc ii}]}$  shows a good correlation with $\sigma_{\star}$, in spite of the fact that the gaseous $\sigma$ values are systematically higher than those of  the stars.

\begin{figure}
 \centering
 \includegraphics[scale=0.55]{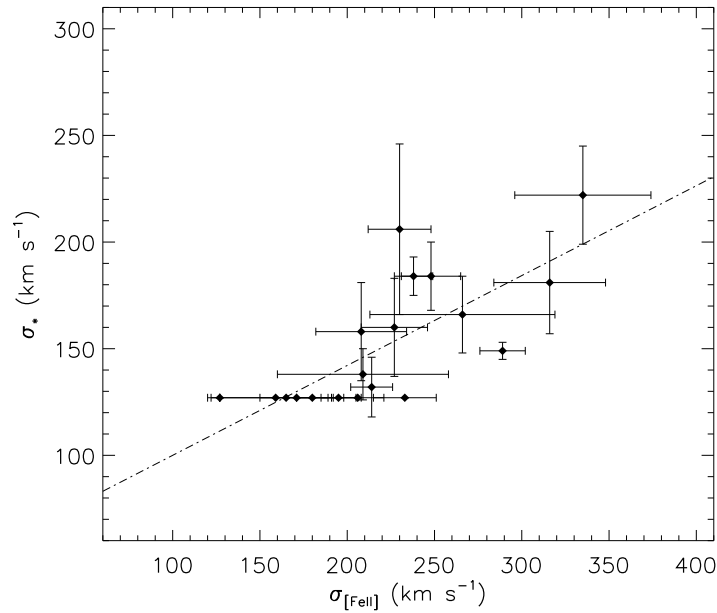}
 \caption{As Fig.~\ref{h2} but for gas velocity dispersion  derived  using [Fe\,{\sc ii}]$\lambda1.257\mu$m for the sub-sample of unbarred galaxies. The dashed line represents the best linear fit of the data, given by: 
$\sigma_\star = 57.9\pm23.5 + (0.42\pm0.10)\times \sigma_{\rm  [Fe\,II]}$.
} 
 \label{fenobar}  
 \end{figure}

As pointed out above, the $M_{\bullet}-\sigma_\star$ relation is calibrated using a parent sample of mostly early-type galaxies. But Table\,\ref{table} shows that about 30\% of the galaxies of our sample are late-type, which can have distinct $\sigma_\star$ values from those of early-type galaxies, since the orbits of the stars in a disk (which dominate in late-type galaxies) are distinct from those in  a bulge (which dominate in early-type galaxies). Additionally, $\approx$\,30\%  of the galaxies of our sample have bars,  and another 30\% have uncertain classifications and are peculiar objects which may not obey the the $M_{\bullet}-\sigma_\star$ relationship.

\citet{xiao11} investigated the $M_{\bullet}-\sigma_\star$ relation for late-type galaxies for a sample of 93 objects with Seyfet 1 nucleus. They examined the  $M_{\bullet}-\sigma_\star$ relationship for subsamples of barred and unbarred host galaxies and found no difference in slope. They only found a mild offset in the relation between low- and high-inclination disk galaxies, with the latter tending to have larger $\sigma_\star$ for a given value of the black hole mass.

The $M_{\bullet}-\sigma_\star$ relationship for galaxies of different Hubble types have also been studied by \citet{graham11} using a sample of 64 galaxies. They found that restricting the sample only to elliptical galaxies, or only to non-barred galaxies result in tighter relations (with less scatter) and a smaller slope than when using the full sample of galaxies. The  $M_{\bullet}-\sigma_\star$ relation obtained when the sample is restricted to barred galaxies only lies $\approx$\,0.45 dex bellow the relation obtained for elliptical and non-barred galaxies.

In order to investigate the effect of the presence of a bar in the $M_{\bullet}-\sigma_{\rm [Fe\,II]}$ relation of Fig.~\ref{fe}, we divided our sample into two sub-samples: one composed of barred galaxies only and the other of unbarred galaxies. We found a much better correlation between $\sigma_{\rm  [Fe\,II]}$ and $\sigma_\star$ for the unbarred galaxies than for the total sample,  as illustrated in Fig.\ref{fenobar}. The correlation coefficient is $R=0.80$ and a linear regression to the relation is given by 

\begin{equation} \label{msigfe_ub}
\sigma_\star = 57.9\pm23.5 + (0.42\pm0.10)\times \sigma_{\rm  [Fe\,II]}.
\end{equation}
On the other hand, no correlation was found for the barred galaxies, for which the correlation coefficient between $\sigma_{\rm  [Fe\,II]}$ and $\sigma_\star$  is only $R=0.20$. 

Approximately half of our sample is composed by early-type galaxies (half of which are barred), the remainder being of late-type galaxies and galaxies with uncertain morphology (usually because they are distant and compact). But the number of galaxies of our sample is not large enough to further investigate the effect of morphology (besides the effect of bars discussed above) on the $\sigma_{\star}-\sigma_{\rm [Fe\,II]}$ relation.  This will be possible only when more near-IR AGN spectra become available, and the equations~\ref{msigfe} and \ref{msigfe_ub}  could  be better calibrated using large non-biased samples of galaxies. 

On the other hand, when applying the $\sigma_{\star}-\sigma_{\rm [Fe\,II]}$ relation to distant galaxies, it will be hard to ascertain a Hubble type to such galaxies, and it will probably better to just use the relation for the whole sample (equation~\ref{msigfe}), in spite of the fact that a relation for restricted Hubble types (e.g., for galaxies without bars) shows less scatter.

Many present/future missions are discovering/will discover large numbers of obscured AGN, such as  the Wise mission \\
($http://www.nasa.gov/mission\_pages/WISE/news/wise20120829.html$) or the Vista Variables in the Via Lactea  survey \\
($http://mwm.astro.puc.cl/mw/index.php/Main\_Page$), which are potential samples to benefit from the relation we have found in order to obtain the SMBH masses, if [Fe\,{\sc ii}] emission lines are present in the spectra.

\subsection{Estimating SMBH masses using $\sigma_{\rm [Fe\,II]}$}

\begin{figure}
 \centering
 \includegraphics[scale=0.6,angle=-90]{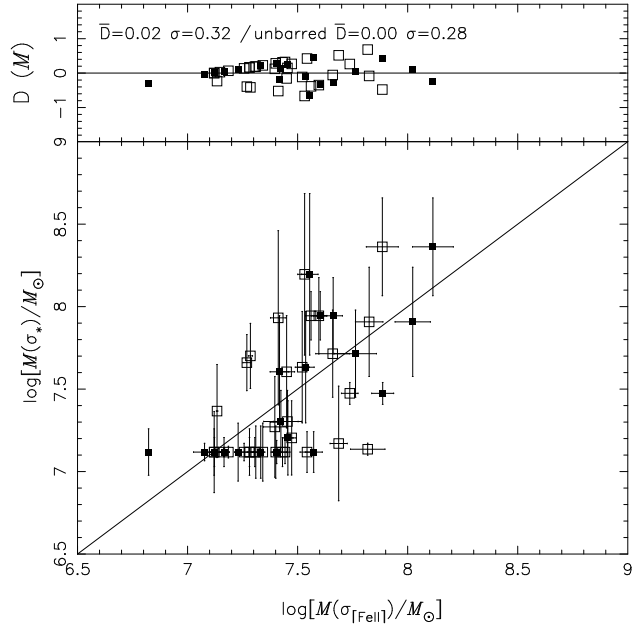}
\caption{Comparison between SMBH mass values obtained directly from the relation with $\sigma_\star$ (y-axis) and  using $\sigma_{\rm  [Fe\,II]}$ (x-axis) to obtain $\sigma_\star$. The filled squares are for the complete sample, for which the eq.\,\ref{msigfe} was used and the opened squares points of for unbarred galaxies, for which the eq.\,\ref{msigfe_ub} was used. The error bars shown include the uncertainties in the calibration of the $M_\bullet - \sigma_\star$ relation.} 
 \label{mbh}  
 \end{figure}

We now evaluate the effect of the scatter introduced by the use of eq.\,\ref{msigfe} in the value of SMBH masses obtained via the  $M_\bullet-\sigma_\star$ relation, which result in higher uncertainties in $M_\bullet$ than those obtained  by using $\sigma_\star$ directly. In order to do this, we use the relation below from \citet{graham11}:

\begin{equation}
{\rm log}(M_{\bullet}/M_{\odot}) = (8.13\pm0.05)+(5.13\pm0.34){\rm log}[\sigma_\star/200\,{\rm km\,s^{-1}}].
\end{equation}
to obtain $M_{\bullet}/M_{\odot}$, using for  $\sigma_\star$ first its measurement from the stellar kinematics and then the value derived from  $\sigma_{\rm  [Fe\,II]}$ using eq.\,\ref{msigfe}. We compare the two values  in Figure~\ref{mbh} (${\rm log}\,M(\sigma_\star)$ vs. ${\rm log}\,M(\sigma_{\rm [FeII]})$), which shows a good agreement between them, with an average difference of $\Delta{\rm log} M_\bullet = {\rm log}M(\sigma_{\rm [Fe\,II})- {\rm log}M(\sigma_\star) = 0.02\pm0.32$.  In the same figure, we show also the masses for the SMBH obtained for the sub-sample of unbarred galaxies as open symbols, using the eq.\,\ref{msigfe_ub} to obtain $\sigma_\star$. There is no diference in the  average value of the SMBH masses obtained, and the mean scatter is 0.28 dex, which is somewhat smaller than the one for the complete sample.

\section{Conclusions}

We have used near-infrared spectroscopic data of a sample of 47 active galaxies in order to investigate possible correlations between the stellar velocity dispersion ($\sigma_\star$), obtained from the fit of the K-band CO absorption band-heads, and the gas velocity dispersion ($\sigma$) obtained from the fit of the profiles of the [S\,{\sc iii}]$\lambda0.95332\mu$m, [Fe\,II]$\lambda1.25702\mu$m, [Fe\,{\sc ii}]$\lambda1.644\mu$m and H$_2\lambda2.12182\mu$m emission lines. The main conclusions of the present paper are:

\begin{itemize}
 
\item Very weak correlations are found between $\sigma_\star$ and both $\sigma_{\rm H_2}$  and $\sigma_{\rm [SIII]}$; %, with correlation coefficients $R=0.35$ and $R=0.32$, respectively.

\item The best correlation is found for the [Fe\,II] emitting gas with $R=0.58$ for the Spearman rank correlation coefficient between $\sigma_\star$ and $\sigma_{\rm  [Fe\,II]}$ (both the $\lambda1.257\mu$m and $\lambda1.644\mu$m emission lines can be used).  A better correlation is found if we exclude the barred galaxies from the sample (R=0.80), while no correlation is found for the sub-sample of barred galaxies.

\item $\sigma_{\rm  [Fe\,II]}$ can thus be used to estimate $\sigma_\star$ for objects for which the stellar velocity dispersion cannot be measured or is unknown.  The best fit of the data is given  by the equation $\sigma_\star = 95.4\pm16.1 + (0.25\pm0.08)\times \sigma_{\rm  [Fe\,II]}$ for the complete sample and $\sigma_\star = 57.9\pm23.5 + (0.42\pm0.10)\times \sigma_{\rm  [Fe\,II]}$ for the sub-sample of unbarred galaxies. 

\item The equations above should be improved and re-calibrated when larger and non-biased samples of  near-IR spectra of AGN  become available.

\item The scatter from a one-to-one relationship between $\sigma_\star$ and its value derived from $\sigma_{\rm [Fe\,II]}$ using the equation above for our sample is 0.07 dex, which is smaller than the scatter of previous relations using $\sigma_{\rm [O\,III]}$ in the optical as a proxy for $\sigma_\star$.

\item The use of $\sigma_{\rm [Fe\,II]}$ in the near-IR instead of $\sigma_{\rm [O\,III]}$ in the optical is particularly important for cases in which the optical spectra  is not available or is obscured, as is the case of many AGN.

\item The comparison of the masses for SMBHs obtained from the direct use of $\sigma_\star$ in the $M_\bullet-\sigma_\star$ relation with those using  $\sigma_{\rm  [Fe\,II]}$ to obtain $\sigma_\star$ reveals only a  small average difference of  $\Delta{\rm log} M_\bullet =0.02\pm0.32$ for the complete sample and $\Delta{\rm log} M_\bullet =0.00\pm0.28$ excluding barred galaxies from the sample.

\end{itemize}

\section*{Acknowledgements}

We thank an anonymous referee for useful suggestions which helped to improve the paper. This paper is based on observations obtained at the Infrared Telescope Facility, which is operated by the University of Hawaii under Cooperative Agreement no. NNX-08AE38A with the National Aeronautics and Space Administration, Science Mission Directorate, Planetary Astronomy Program. This research has made use of the NASA/IPAC Extragalactic Database (NED) which is operated by the Jet Propulsion Laboratory, California Institute of Technology, under contract with the National Aeronautics and Space Administration. We acknowledge the usage of the HyperLeda database (http://leda.univ-lyon1.fr). This work has been partially supported by the Brazilian institution CNPq and FAPERGS.

\label{lastpage}

\end{document}